\documentclass{article}

\usepackage{PRIMEarxiv}

\usepackage[utf8]{inputenc} 
\usepackage[T1]{fontenc}    
\usepackage{hyperref}       
\usepackage{url}            
\usepackage{booktabs}       
\usepackage{amsfonts}       
\usepackage{nicefrac}       
\usepackage{microtype}      
\usepackage{lipsum}
\usepackage{fancyhdr}       
\usepackage{graphicx}       
\graphicspath{{media/}}     

\usepackage{amssymb}
\usepackage{amsmath}

\title{Improving on the Markov-Switching Regression Model by the Use of an Adaptive Moving Average
}

\author{
  Piotr Pomorski, Denise Gorse \\
  Department of Computer Science \\
  University College London \\
  London\\
  \texttt{\{P.Pomorski, D.Gorse\}@cs.ucl.ac.uk} \\
}

\begin{document}
\maketitle

\begin{abstract}
Regime detection is vital for the effective operation of trading and investment strategies. However, the most popular means of doing this, the two-state Markov-switching regression model (MSR), is not an optimal solution, as two volatility states do not fully capture the complexity of the market. Past attempts to extend this model to a multi-state MSR have proved unstable, potentially expensive in terms of trading costs, and can only divide the market into states with varying levels of volatility, which is not the only aspect of market dynamics relevant to trading. We demonstrate it is possible and valuable to instead segment the market into more than two states not on the basis of volatility alone, but on a combined basis of volatility and trend, by combining the two-state MSR with an adaptive moving average. A realistic trading framework is used to demonstrate that using two selected states from the four thus generated leads to better trading performance than traditional benchmarks, including the two-state MSR. In addition, the proposed model could serve as a label generator for machine learning tasks used in predicting financial regimes ex ante.
\end{abstract}


\section{Introduction}
Financial markets are characterised by periods of evolving, low-volatility growth separated with a disruptive, high-volatility contractions. These regimes can be distinguished by significant changes in asset returns, variances, and correlations, laying a groundwork for accurate detection techniques to be exploited by portfolio managers. One such detection technique is the \emph{Markov-switching regression model} (MSR), introduced by Goldfeld \& Quandt in 1973 \cite{ref1}, and later extended by Hamilton in 1989 \cite{ref2} and Krolzig in 1997 \cite{ref3}, which has since become one of the most popular statistical methods to distinguish regime shifts in economics and finance. However, the original two-state MSR model is not ideal, as two volatility states are insufficient to describe the complexities of market dynamics, and attempts to use a three- or higher state MSR can suffer from instability due to overly frequent regime switches \cite{ref4}, which in addition adversely impact portfolio returns by increasing the costs of trading. Furthermore, volatility by itself is not an infallible indicator of up- or down-trending markets. This paper improves on the current state of the art by combining the two-state MSR with the use of \emph{Kaufman's adaptive moving average} (KAMA) \cite{ref5}, a method that is both stable (avoiding spurious regime shift detections) and accurate in locating the times of onset of regime switches. This combination generates four detected regimes based not on volatility alone, allowing, ultimately, a focus on those two market states with the greatest trading value, namely low-variance bullish and high-variance bearish, with trading results that will demonstrate the utility of the method.

\section{Background and Related Work}
\label{sec:back_and_rel}


\subsection{Two-State Markov-Switching Regression (MSR) Model}
The model to be described here is also known as the \emph{two-state Markov-switching dynamic regression model} \cite{ref3} and will be used throughout this paper as a fundamental part of the proposed regime-switching model, as well as (in its unenhanced version) one of the models used for performance comparison.

The key to the MSR model’s widespread use is its ability to incorporate observed characteristics of asset returns, such as asymmetries, autocorrelation, and volatility clustering \cite{ref6}. This gives a way to separate the underlying process $y_t$  into $S_t$  states ('regimes') at times $t \in \{0,1,…,T\}$, where $S_t \in \{0,1,…,k\}$. Assuming the process $y_t$  refers to daily log returns $\ln r_t$, the model can be specified as

\begin{equation}
\ln r_t = \mu_{S_t} + \ln r_{t-1}\beta_{S_t} + \sigma_{S_t} \epsilon_t, \ \ \epsilon_t \sim N(0,1),
\end{equation}

where $\mu_{S_t}$ is a state-dependent intercept, $\beta_{S_t}$ is a state-dependent coefficient of lagged log returns, and $\sigma_{S_t}$ is a state-dependent volatility \cite{ref3}. With $S_t \in \{0,1\}$, the governing dynamics of the underlying regime $S_t$ are considered to follow a time-homogeneous Markov chain with fixed transition probabilities $p,q \in [0,1]$,

\[
\begin{pmatrix}
p & {1-p}\\
{1-q} & q
\end{pmatrix}
\]

where

\begin{equation}
p = Pr(S_t=0 | S_{t-1}=0), \ \ q = Pr(S_t=1 | S_{t-1}=1).
\end{equation}

If $(\delta,1 -\delta)$, with $\delta = Pr(S_t = 0) \in [0,1]$, is the initial distribution of the Markov chain, then the model is completely specified by the vector of parameters
$\theta = (p, q, \mu_0, \mu_1, \beta_0, \beta_1, \sigma_0, \sigma_1, \delta)$ \cite{ref4}, and the estimation of $\theta$ is performed using Gibbs sampling, a type of Markov chain Monte Carlo algorithm described in \cite{ref7}.

If $\Hat{\theta}_t$ denotes the estimate of $\theta_t$ , then for the two volatility states $i \in \{0,1\}$ the probabilities at time $t \in \{0,1,…,T\}$  are given by

\begin{equation}
p_t^i = Pr(S_t = i \ | \ \ln r_t ; \Hat{\theta}_t ),
\end{equation}

in which $S_t = 0$ and $S_t = 0$ are the low and high volatility periods, respectively. $p_t^i = Pr(S_t = i \ | \ \ln r_t ; \Hat{\theta}_t )$ estimates the probability of a low ($i = 0$)  or high ($i = 1$)  volatility regime at time $t$; the higher the $p_t^i$, the more likely the process is persistent \cite{ref4}.

\subsection{Kaufman's Adaptive Moving Average (KAMA)}
The popularity of moving averages in trading strategies has prompted the creation of many different forms of these trend-following indicators. Some of these, such as simple moving averages, are better at smoothing the underlying price, and some, such as the zero-lag moving average, are better at minimising the lag, though these objectives cannot be achieved simultaneously. An adaptive moving average is, however, a specific class of moving average that attempts to do so, accounting for both trend and volatility factors to balance the smooth-lag issue \cite{ref9}. (Note: From this point on, this paper will use the terms 'adaptive moving average', 'Kaufman’s adaptive moving average', and the abbreviation 'KAMA' interchangeably.)

The calculation of KAMA begins with determining an \emph{efficiency ratio} (ER). The ER helps KAMA identify and adapt to ever-changing asset conditions by measuring the relative speed of price movement from one period to another. Assuming a daily frequency over time $t \in \{0,1,…,T\}$ , the ER is defined by

\begin{equation}
E R_t = \frac{M_t}{V_t},
\end{equation}

where

\begin{equation}
M_t = P_t - P_{t-n}, \ \ V_t = \sum _{i=1}^{n} | P_t - P_{t-1} |,
\end{equation}

and the momentum, $M_t$, is the change in closing price $P$ over a period of length $n$ ($n$-period) and the volatility, $V_t$, is the sum of the absolute value of daily closing price changes during an $n$-period. The efficiency ratio is constrained to $0 \leq ER \leq 1$, such that a significant price change as a proportion of low volatility will bring ER closer to 1, whereas a minor price change as a proportion of high volatility will yield an ER closer to 0. In other words, values close to 1 indicate a clearly defined trend in the price, whereas values nearing 0 indicate a consolidating and directionless market.

The ER is embedded into KAMA’s full formula as a part of its scaled smoothing coefficient $C$. Considering again the daily frequency over time $t \in \{0,1,…,T\}$, the adaptive moving average is computed as

\begin{equation}
KAMA_t = KAMA_{t-1}  + C_t  (P_t -KAMA_{t-1} ),
\end{equation}

where the scaled smoothing coefficient $C$ over time $t$ is expressed as

\begin{equation}
C_t = [ER_t  (k_s - k_l) + k_s ]^2 .
\end{equation}

$k_s$ and $k_l$ in the above are smoothing constants relevant to a pair of different (short-term and long-term) simple moving averages over $n$ period, calculated as

\begin{equation}
k_s = \frac{2}{n_s + 1}, \ \  k_l = \frac{2}{n_l + 1},
\end{equation}

where $n_s$  and $n_l$  refer to shorter and longer time windows, respectively \cite{ref5}.

By combining the trending aspect of moving averages and the volatility factor of the ER, KAMA is able to both identify the global price trend and to accurately locate larger turning points. Thus, the potential trader would not need to so frequently switch position by buying and selling over local price swings \cite{ref5}. MSR models also seek to determine the current state of the process with highest probability, in order to avoid switching states too often. However, the main difference is that KAMA is primarily used to capture the global trend, bullish or bearish, whereas an MSR model detects disruptions in the variance of the underlying series.

\subsection{Related Work}

Since they were first introduced by \cite{ref1}, MSR models have been discussed broadly in economic literature. Their performance has been analysed on macroeconomic variables, such as GDP \cite{ref2} and inflation \cite{ref8}, and they have been additionally applied to financial time series, such as equities \cite{ref5} and foreign exchange \cite{ref9}.

The original, two-state MSR model is, however, not optimal, for reasons discussed in the Introduction. However, as also discussed previously, extension to three-state MSR models comes at a cost, since they are deemed unstable due to their tendency to shift regimes too frequently \cite{ref4}. There are also concerns about potential over-fitting, as beyond the two-state model it is necessary to pre-estimate the optimal number of states, applying each version of the MSR model to each asset in order to find the optimal number of states \cite{ref10}, and hoping that the same optimal number of regimes will continue in the future. Finally, these multi-state MSR models do not solve the issue of overly frequent switches—on the contrary, there is a high potential of increasing them.

However, there is another potential solution to finding a better partition of the market dynamics \cite{ref11}, which is to overlay the two-state model with moving averages, which method allows partitioning without frequent state switches, resulting in a four-state model in line with classic Wyckoff theory \cite{ref12}. But although the model of \cite{ref11} has been found to add value in equity strategies, it is constrained by the data required to calculate a necessary technical indicator, which requires period-high and -low prices. This limits application of the model to assets with a recorded history of high and low prices, even when their closing prices have been available for a much longer time. The work of this paper has taken inspiration from that of \cite{ref11}, replacing the custom Keltner Channels \cite{ref13} in that model with KAMA, which does not require the recording of high and low prices; as will be seen in the Results section, the new model is both highly effective and applicable to a wide range of financial assets.

\section{Data}

Setting up the proposed and benchmark models requires daily closing price data only. There are 56 assets in total, divided into four classes: equities (24 datasets), exchange rates (FX) (13 datasets), commodities (12 datasets), and fixed income (seven datasets). The choice of these assets is based on the extent of their use in the financial industry (for instance, major equity markets for stock indices, G10 countries for FX), to ensure there will be sufficient liquidity to minimise the risk of market manipulation. In addition to the above-mentioned assets, a cash index is also used in the trading strategy detailed later in the Methodology section. The rationale behind the use here of a cash index stems from its low volatility and ever-rising trend. With the rare exception of periods of negative interest rates, such as after the financial crisis of 2008-2009, it is nearly impossible to lose whilst betting on cash in the longer term, which makes it a useful 'money-parking' tool. 
The start dates for the datasets vary, though are usually in the 1980s-1990s. The end dates are 26/03/2021 for equities, foreign exchange, commodities, and the cash index, and 08/01/2021 for fixed income. For each asset, that 85 \% of the data closest to the start date is used for optimisation of the KAMA model, using procedures to be described below, while that 15 \% of the data closest to the end date is used for out of sample testing.

\section{Methodology}

This section is constructed as follows: first, the process of combining the two-state MSR model with KAMA is described. Second, the optimisation of the KAMA component within the combined model is outlined. Third, benchmark models used for performance comparison are briefly discussed, and finally the trading strategy, which assesses the tested models’ abilities to separate regimes, is described.

\subsection{Combining the Two-State MSR Model with KAMA}

The proposed regime-switching model is initiated with a two-state MSR model to detect high and low variance periods for each selected asset. Considering states $S_t \in \{0,1\}$, and the 50 \% level as a cut-off point for smoothed probabilities in the MSR model, this initial phase of model construction results in the separation:

\begin{itemize}
  \item Low variance regimes, defined as ones where the filtered probability of state $S_t=0$ is higher than 50 \%.
  \item High variance regimes, defined as ones where the filtered probability of state $S_t=1$ is higher than 50 \%.
\end{itemize}

KAMA then works as an overlay to divide these low and high volatility periods into bullish and bearish regimes. However, its practical application also requires a mechanism to generate trading decisions, and for this purpose Kaufman's adaptive moving average is embedded within a construction called the filter which generates a signal to enter or exit positions of interest. Over a $t$-day period, the filter $f$ is computed as $f_t = \gamma \sigma (KAMA_t)$, where

\begin{equation}
\sigma (KAMA_t) = \sqrt{({\sum _{t=1}^{n} x_t^2} - \frac{{\sum _{t=1}^{n} x_t}}{n})} \ ,
\end{equation}

is the standard deviation of the change in KAMA over $n$ days, where $n \leq t$ , $x_t = KAMA_t - KAMA_{t-1}$, and the parameter $\gamma$ is as discussed in the Background section.

Calculating both KAMA and the filter allows the construction of a strategy for trading in both bullish and bearish regimes:

\begin{itemize}
  \item Bullish, hence buy, when KAMA advances above its low over a prior period of $n$ days by a value greater than the filter.
  \item Bearish, hence sell, when KAMA descends below its low over a prior period of $n$ days by a value greater than the filter
\end{itemize}

Combining the above with the MSR model’s results, the proposed regime-switching model, which we will refer to as the \emph{KAMA+MSR} model, separates the underlying price series into four regimes:

\begin{itemize}
  \item Low variance and bullish when the filtered probability of state $S_t = 0$ is higher than 50 \%, and KAMA rises above its low over a prior period of $n$ days by a value greater than the filter.
  \item Low variance and bearish when the filtered probability of state $S_t = 0$ is higher than 50 \%, and KAMA falls below its low over a prior period of $n$ days by a value greater than the filter.
  \item High variance and bullish when the filtered probability of state $S_t = 1$ is higher than 50 \%, and KAMA rises above its low over a prior period of $n$ days by a value greater than the filter.
  \item High variance and bearish when the filtered probability of state $S_t = 1$ is higher than 50 \%, and KAMA falls below its low over a prior period of $n$ days by a value greater than the filter.
\end{itemize}

However, while all four regimes are of academic interest, not all are equally valuable for trading. The second and third are likely to be less useful because in the first of these cases the potential for profit is low, while in the second the risk is overly high. In preliminary experiments this was confirmed by the low Sharpe ratios obtainable during these periods. Thus, in the trading results to be reported in Section 5, active trading is carried out only in the first and fourth of the above-listed regimes.

\subsection{Optimisation of the KAMA Component of the Model}

The MSR component of the KAMA+MSR model does not require optimisation, as it is parameter-free. In contrast, the KAMA model requires the optimisation, over the training period that comprises the initial 85 \% of the data, of $\theta_h = (n, n_s, n_l, \gamma)$, that vector of parameters discussed in the Background section, where $n$ is a moving window for the efficiency ratio ER and the filter $f$, $n_s$ is a moving window for a short-term smoothing constant $k_s$, $n_l$ is a moving window for a long-term smoothing constant $k_l$, and $\gamma$ is the control parameter in the filter $f$  term.

Beginning with the initial vector of parameters $\theta_h$, the algorithm constructs the proposed regime-switching model by overlaying the calculated KAMA on a two-state Markov-switching dynamic regression. Based on the outcome of this initial combination, the model splits each asset price into multiple periods ('segments') which are classified as being in one of the four regimes. Subsequently, the price slope (bullish vs. bearish) and log returns volatility (low vs. high variance) of each segment is computed, so that the optimised model’s accuracy can be inspected. The parameter values are then adjusted to optimise performance during the training period.

The test of meaningful separation of regimes is done using a method based on K-Means clustering \cite{ref14}, with a two-step walk-forward cross-validation and a custom scoring function. In order to implement cross-validation, the regime slope and volatility data is split into non-shuffled training and validation sets, whereby training is 50 \% and 75 \% of the training dataset in the first and second steps, respectively, and the consecutive 25 \% of the training data are used for validation in both steps. The clustering ability of K-Means is employed on the sliced training sets to investigate whether the already refined regimes differ by slopes and volatilities. Given that this will be a four-state model, K-Means here assumes four clusters; the clustering algorithm additionally uses ten random initialisations of centroids, the same random seed for each model, and 300 maximum iterations per initialisation. The K-Means method then predicts clusters for the validation sets, and the generated predictions are contrasted with the model's initially chosen regime labels.

To achieve this, the KAMA parameter optimisation algorithm uses a custom function called the \emph{misclassification score}. This scoring function groups clusters and regime labels into a $4 \times 4$ matrix in order to analyse the dominant number of detected segments (periods of time that fall into a given class) within each row and column, as in Table~\ref{table1}, and perform appropriate actions.

\begin{table}[]
\caption{Example grouping matrix; matrix entries denote the number of detected segments}
\centering
\begin{tabular}{|c|c|c|c|c|}
\hline
\textbf{Cluster} & \textbf{\begin{tabular}[c]{@{}c@{}}Low variance /\\ bullish\end{tabular}} & \textbf{\begin{tabular}[c]{@{}c@{}}Low variance /\\ bearish\end{tabular}} & \textbf{\begin{tabular}[c]{@{}c@{}}High variance /\\ bullish\end{tabular}} & \textbf{\begin{tabular}[c]{@{}c@{}}High variance /\\ bearish\end{tabular}} \\ \hline
Cluster 1        & 1                                                                         & 0                                                                         & 0                                                                          & 0                                                                          \\ \hline
Cluster 2        & 0                                                                         & 5                                                                         & 1                                                                          & 9                                                                          \\ \hline
Cluster 3        & 0                                                                         & 0                                                                         & 2                                                                          & 0                                                                          \\ \hline
Cluster 4        & 3                                                                         & 0                                                                         & 3                                                                          & 0                                                                          \\ \hline
\end{tabular}
\label{table1}
\end{table}

The misclassification score initially loops over rows to subtract the dominant number of segments from the row sum. Thus, in the example, in Clusters 1 and 3, since there is only one regime detected, the score is 0, which indicates lack of misclassification within this cluster. However, in Cluster 2, there are 6 segments (out of 15) whose underlying slope and volatility characteristics do not resemble the dominant label, high variance and bearish, so the score is 6. In Cluster 4, there are two dominant labels, each mismatched (with respect to the other) with a score of 3, and we would in a case like this assign an overall score for the cluster of 3. The number of misclassifications per cluster is then summed up to indicate how many segments have been inaccurately labelled by the optimised model. The total number is finally divided by the length of the predicted dataset to form a ratio that can be compared between optimisation trials. Ideally, the final misclassification score should equal 0, representing a perfect alignment between the K-Means discovered clusters and the proposed regime-switching model, though this degree of alignment is unlikely in practice and hence the optimisation algorithm is run 50 times to find the best-available vector of parameters $\theta_h$ for the training period, which are saved for use in the test period.

\subsection{Benchmark Models}

The most obvious candidate as a benchmark is the two-state MSR model, being a component of the proposed KAMA+MSR model; comparison of MSR with the proposed model will reveal the value of the KAMA component. Initial experimentation determined that the model of \cite{ref11}, developed for equities, was not effective over the full range of asset classes considered in this work. In addition, this model requires knowledge of price highs and lows, and this data was not available for the full desired training period, that stretched back into the 1980s-1990s. This model was thus set aside. Additionally, while a three-state MSR model \cite{ref15} was considered, it was found to be unstable (as discussed previously, multi-state MSR models typically have this associated difficulty) but to be effective when its three states were reduced to two, by removing that state which had medium variance, and it was in this form included in the set of benchmark models. It should be noted that the two chosen benchmark models, like the MSR component of the proposed model, are parameter-free and hence do not need a training period parameter optimisation.

\subsection{Implementing a Trading Strategy for Performance Testing}

The trading strategy requires an allocation of weights (adding to 100 \%) between the asset and cash. These are optimised over the training period (earliest 85 \% of the data, for each asset), by comparing the effectiveness of 1000 random weight allocations. Both annualised returns and the adjusted Sharpe ratio (ASR) \cite{ref16}

\begin{equation}
ASR_t = MAR_t / \sigma^{\frac{MAR_t}{abs(MAR_t)}}
\end{equation}

are calculated, on a per-segment basis. The adjusted Sharpe ratio is a valuable measure because it penalises negative variance especially strongly, while treating positive returns in the same manner as the standard Sharpe ratio. $MAR_t$ in the above denotes mean excess annualised return over $t$ days, $\sigma$  is the annualised volatility, and $abs(MAR_t)$ denotes mean excess annualised return over $t$ days calculated from only positive excess returns. In this work, mean excess annualised returns equal mean annualised returns, as for simplification the risk-free rate is set to 0.

Inevitably, the asset returns are diminished by two-way transaction costs, i.e., ones incurred during both buying and selling. Table~\ref{table2} lists the costs assumed in this work, as given in \cite{ref17} (these values being considered in the industry to be conservative and useful estimates of costs):

\begin{table}[]
\caption {Two-way (buy, sell) trading costs for each asset class}
\centering
\begin{tabular}{|l|c|c|c|c|}
\hline
\multicolumn{1}{|c|}{\textbf{Asset class}} & \textbf{\begin{tabular}[c]{@{}c@{}}Brokerage  \\ commissions (\%)\end{tabular}} & \textbf{\begin{tabular}[c]{@{}c@{}}Bid-ask\\ spread (\%)\end{tabular}} & \textbf{\begin{tabular}[c]{@{}c@{}}Market\\ impact (\%)\end{tabular}} & \textbf{\begin{tabular}[c]{@{}c@{}}Total \\ cost (\%)\end{tabular}} \\ \hline
Equities                                   & 0.14                                                                            & 0.13                                                                   & 0.53                                                                  & 0.8                                                                 \\ \hline
Currencies                                 & 0                                                                               & 0.13                                                                   & 0                                                                     & 0.13                                                                \\ \hline
Commodities                                & 0.14                                                                            & 0.13                                                                   & 0                                                                     & 0.27                                                                \\ \hline
Fixed income (ETFs)                        & 0.14                                                                            & 0.13                                                                   & 0.53                                                                  & 0.8                                                                 \\ \hline
\end{tabular}
\label {table2}
\end{table}

The differing costs between the classes of assets stem from the sizes of these markets and the potential influence a trader may have on them. For instance, it is difficult to have any significant impact on foreign exchange, unless the trader is an intervening central bank. Additionally, since in this research raw currencies are traded, there are no brokerage commissions, and the only cost remaining is the bid-ask spread. In the case of commodities, the trader must account for brokerage fees, as well as the bid-ask spread on futures contracts. However, only on rare occasions would the trade be large enough to significantly move the entire market. For equity indices and fixed income ETFs, the full cost is taken into consideration, since on top of commissions and the bid-ask spread there is a potential market impact.

The KAMA+MS-DR model is finally compared to the benchmark models, in relation to both annualised returns and ASR, via a 'winning score ratio', $WS$, calculated as

\begin{equation}
WS = \frac{(Z_{winner} - Z_{runner-up})}{Z_{winner}}
\end{equation}

where $Z_{winner}$ refers to the model with the highest annualised returns or adjusted Sharpe ratio, and $Z_{runner-up}$ refers to the second-best method. By computing the ratio in the above equation, it is possible not only to indicate the winner, but also by how much better it is relative to its closest opponent. After the winning score ratio has been obtained for both weighted annualised returns and the adjusted Sharpe ratio the two resulting ratios are combined as an average to reveal the model with the most effective risk-reward approach to detecting financial regimes. This combination is done as an equally weighted average; depending on an investor's preference either returns or Sharpe ratio could be more highly weighted, but a ratio of 1:1 is the most reasonable assumption in the absence of such a stated preference.

\section{Results}

The results of using the KAMA+MSR and two benchmark regime detection models within the out-of-sample test periods (last 15 \% of the data, for each asset considered) are presented in Table~\ref{table3}. It is clear that the proposed model performs strongly, in each of the two scoring metrics, for equities and fixed income. In the case of equities, the notably higher adjusted Sharpe ratio for KAMA+MSR points at a good balance between returns and volatility, which suggests a solid blend between the use of trend in predicting regime switches within KAMA and the use of variance for this purpose in Markov-switching regression. It is also worth pointing out that the proposed model achieves such results regardless of the equity region. Whether it is the USA, Europe, or Emerging Markets, the KAMA+MSR model seems to be resistant to both global and local shocks (though more indices, in particular from the Emerging Markets region, could strengthen this argument).
Results for commodities and foreign exchange are more mixed, as can be seen from the table. However, after combining the two winning scores the KAMA+MSR model can be seen to have the overall best performance of those models tested, with an especially strong result, again, for equities and fixed income.

\begin{table}[]
\caption{Out-of-sample trading performance results, in which \emph{MSR 2S} refers to the two-state MSR model, \emph{MSR 3S—>2S} to the three-state MSR model converted to two-state form, and \emph{KAMA+MSR} to the regime detection model proposed in this work}
\centering
\begin{tabular}{|lccc|}
\hline
\multicolumn{4}{|c|}{\textbf{Average Winning Score Ratio: Highest Adjusted Sharpe Ratio}}                                                             \\ \hline
\multicolumn{1}{|l|}{}             & \multicolumn{1}{c|}{\textbf{MSR 2S}} & \multicolumn{1}{c|}{\textbf{MSR 3S—\textgreater{}2S}} & \textbf{KAMA+MSR} \\ \hline
\multicolumn{1}{|l|}{Equities}     & \multicolumn{1}{c|}{0.03}            & \multicolumn{1}{c|}{0.03}                             & 0.33              \\ \hline
\multicolumn{1}{|l|}{Commodities}  & \multicolumn{1}{c|}{0.25}            & \multicolumn{1}{c|}{0.24}                             & 0.13              \\ \hline
\multicolumn{1}{|l|}{FX}           & \multicolumn{1}{c|}{0.09}            & \multicolumn{1}{c|}{0.29}                             & 0.15              \\ \hline
\multicolumn{1}{|l|}{Fixed Income} & \multicolumn{1}{c|}{0.06}            & \multicolumn{1}{c|}{0.01}                             & 0.22              \\ \hline
\multicolumn{4}{|c|}{\textbf{Average Winning Score Ratio: Highest Weighted Annual Returns}}                                                           \\ \hline
\multicolumn{1}{|l|}{}             & \multicolumn{1}{c|}{\textbf{MSR 2S}} & \multicolumn{1}{c|}{\textbf{MSR 3S—\textgreater{}2S}} & \textbf{KAMA+MSR} \\ \hline
\multicolumn{1}{|l|}{Equities}     & \multicolumn{1}{c|}{0.08}            & \multicolumn{1}{c|}{0.03}                             & 0.10              \\ \hline
\multicolumn{1}{|l|}{Commodities}  & \multicolumn{1}{c|}{0.02}            & \multicolumn{1}{c|}{0.12}                             & 0.27              \\ \hline
\multicolumn{1}{|l|}{FX}           & \multicolumn{1}{c|}{0.22}            & \multicolumn{1}{c|}{0.02}                             & 0.19              \\ \hline
\multicolumn{1}{|l|}{Fixed Income} & \multicolumn{1}{c|}{0.25}            & \multicolumn{1}{c|}{0.00}                             & 0.18              \\ \hline
\multicolumn{4}{|c|}{\textbf{Combined Average Winning Score Ratio}}                                                                                   \\ \hline
\multicolumn{1}{|l|}{}             & \multicolumn{1}{c|}{\textbf{MSR 2S}} & \multicolumn{1}{c|}{\textbf{MSR 3S—\textgreater{}2S}} & \textbf{KAMA+MSR} \\ \hline
\multicolumn{1}{|l|}{Equities}     & \multicolumn{1}{c|}{0.11}            & \multicolumn{1}{c|}{0.06}                             & 0.43              \\ \hline
\multicolumn{1}{|l|}{Commodities}  & \multicolumn{1}{c|}{0.27}            & \multicolumn{1}{c|}{0.35}                             & 0.40              \\ \hline
\multicolumn{1}{|l|}{FX}           & \multicolumn{1}{c|}{0.31}            & \multicolumn{1}{c|}{0.31}                             & 0.35              \\ \hline
\multicolumn{1}{|l|}{Fixed Income} & \multicolumn{1}{c|}{0.31}            & \multicolumn{1}{c|}{0.01}                             & 0.40              \\ \hline
\end{tabular}

\end{table}
\label{table3}
\section{Discussion}

This paper presented an enhancement of the two-state Markov-switching regression model (MSR) by the addition of Kaufman’s adaptive moving average (KAMA), resulting in a new model, which we term the KAMA+MSR model. Unlike multi-state MSR models, which seek only to subdivide variance (e.g., for a three-state MSR, into low, medium, and high variance), and which have proven both unstable and to result in large trading costs, KAMA+MSR seeks to instead identify regimes as combinations of variance and market trend (bull or bear), with an accurate timing of onset of each state, and without overly-frequent regime switches. The proposed model initially discovers four states, of which two combinations (low variance bullish and high variance bearish) were used actively in the trading strategy, with the remaining two (low variance bearish and high variance bullish) being periods in which it was deemed wisest to refrain from action. Due to the use of both variance and trend measures to detect regime switches in the KAMA+MSR model it proved able to outperform, on average, the comparison models in each asset class considered. However, we note that such an outcome could in principle be due to the strategy itself (asset vs. cash), and hence intend to implement alternative strategies in future work.

There are three further ways in which the work of this paper could be extended. First, and most importantly, though the proposed model overall outperformed the benchmark models, this outperformance was less pronounced for foreign exchange and commodities, and it is possible that an alternative to Markov switching could do better in combination with KAMA in these cases. Second, as already mentioned, the current trading strategy discards two of the four discovered regimes as being likely to be unprofitable; however, more complex trading strategies, for example using derivatives-based hedging, could potentially make use of these regimes. Finally, and most importantly for future work, the KAMA+MSR model’s ability to smoothly and accurate distinguish regimes will allow it to be used as a regime label generator for machine learning tasks in which the aim is to predict such regimes ex ante.

\bibliographystyle{unsrt}  
\bibliography{references}

\end{document}